\documentclass{osa-article}

\journal{oe}

\begin{document}

\title{A hybrid integrated InP-Si$_3$N$_4$ diode laser with 40-Hz intrinsic linewidth}

\author{Youwen Fan,\authormark{1} Albert van Rees,\authormark{1} Peter J.M. van der Slot,\authormark{1,*} Jesse Mak,\authormark{1} Ruud M. Oldenbeuving,\authormark{2} Marcel Hoekman,\authormark{2} Dimitri Geskus,\authormark{2} Chris G.H. Roeloffzen,\authormark{2} and Klaus-J. Boller\authormark{1}}

\address{\authormark{1}Laser Physics and Nonlinear Optics, Mesa+ Institute for Nanotechnology, Department for Science and Technology, University of Twente, Enschede, The Netherlands\\
\authormark{2}LioniX International BV, Enschede, The Netherlands}

\email{\authormark{*}Corresponding author: p.j.m.vanderslot@utwente.nl}

\begin{abstract}
We demonstrate a hybrid integrated and widely tunable diode laser with an intrinsic linewidth as narrow as 40~Hz, achieved with a single roundtrip through a low-loss feedback circuit that extends
the cavity length to 0.5 meter on a chip. Employing solely dielectrics for single-roundtrip, single-mode resolved feedback filtering enables linewidth narrowing with increasing laser power, without limitations through nonlinear loss. We achieve single-frequency oscillation with up to 23~mW fiber coupled output power, 70-nm wide spectral coverage in the 1.55 $\mu$m wavelength range with 3~mW output, and obtain more than 60 dB side mode suppression. Such properties and options for further linewidth narrowing render the approach of high interest for direct integration in photonic circuits serving microwave photonics, coherent communications, sensing and metrology with highest resolution.
\end{abstract}

\section{Introduction}
\label{sec:intro}

Semiconductor lasers with narrow linewidth and wide tunability are of central interest in photonic applications where controlling the optical phase is essential, for instance for microwave photonics ~\cite{marpaung_2019NP}, optical beamforming networks~\cite{zhuang_2010JLT}, coherent optical communications~\cite{zhang_2009PTL}, light detection and ranging (LIDAR)~\cite{koroshetz_2005}, optical sensing~\cite{he_2011NT}, or precision metrology and timing, including GPS systems ~\cite{hemmerich_1990OC, jiang_2011NP,newman_2019O}. Of particular interest are narrow linewidth semiconductor lasers for pumping Raman and Brillouin lasers~\cite{spillane_2002N,li_2017O,gundavarapu_2019NPhot}, integration into functional photonic circuits, to serve as light engines, such as for electrically driven and fully integrated Kerr frequency combs~\cite{stern_2018N, raja_2019NatCom}.

A measure for a laser's intrinsic phase stability is the intrinsic linewidth (Schawlow-Townes limit), which can only be narrowed via increasing the photon lifetime of the laser cavity, or via increasing the laser power ~\cite{schawlow_1958PR, lax_1968PR}.  However, in monolithic diode lasers both are problematic due to linear and nonlinear loss. The intrinsic waveguiding loss in these semiconductor amplifiers is high, which limits the photon lifetime. Furthermore, the spectral filtering circuitry required for single-frequency oscillation causes additional loss, while efficient output coupling decreases the lifetime further. Also, at high laser power nonlinear loss occurs. This leads to large intrinsic linewidths typically in the range of a MHz~\cite{akulova_2002JSTQE}.

Many orders of magnitude smaller intrinsic linewidths have been achieved with hybrid and heterogeneously integrated diode lasers, ultimately reaching into the sub-kHz-range~\cite{boller_2019Phot}. In all these approaches the cavity is extended with additional waveguide circuitry fabricated from a different material platform selected for low loss. For extending the cavity length and maintaining single longitudinal mode oscillation, spectral filtering has mostly been based on microring resonators employing Si waveguides~\cite{kita_2015APL,hulme_2013OE,kobayashi_2015JLT, tran_2020JSTQE}, SiON~\cite{matsumoto_2010OFCC}, SiO$_2$~\cite{debregeas_2014ISLC} and Si$_3$N$_4$~\cite{oldenbeuving_2013LPL, fan_2014SPIE, fan_2017CLEO}, thereby reducing the intrinsic linewidth from hundreds of~kilohertz~\cite{hulme_2013OE, fan_2017CLEO} to 220~Hz~\cite{tran_2020JSTQE}. Silicon waveguides bear the advantage of heterogeneous integration~\cite{santis_2018PNAS, huang_2019O}. However, beyond certain intra-cavity intensities and laser powers, using silicon limits the lowest achievable linewidth through nonlinear loss~\cite{vilenchik_2015PROC, santis_2018PNAS}, specifically, due to two-photon absorption across the relatively small bandgap of silicon \cite{kuyken_2017NANOPHOT}. Avoiding high intensities is difficult when having to select a single longitudinal mode within the wide semiconductor gain spectrum, because high-finesse filtering for strong side mode suppression is associated with resonantly enhanced power. Relying on external amplification and operating the diode laser at low power is not a viable route, because the linewidth increases inversely with lowering the laser output~\cite{schawlow_1958PR}.

\begin{figure}[tbp]
	\centering
		\includegraphics[width=0.6\linewidth]{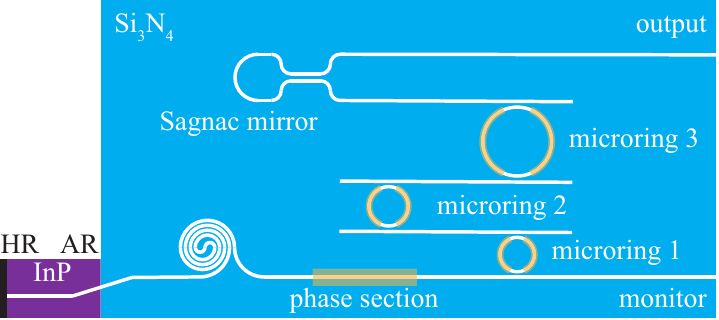}
	\caption{\label{fig:hybrid_laser} Schematic view of the hybrid laser comprising an InP gain section and a Si$_{3}$N$_{4}$ feedback circuit that extends the cavity length physically via a spiral, with a length of 33~mm, and optically via three ring resonators. The cavity mirrors are formed by the HR coating on the back facet of the gain section and the Sagnac mirror. The combined total optical length is significantly larger than the optical length of the solitary semiconductor chip.} 
\end{figure}

To overcome these inherent limitations of the silicon platform, we use a wide bandgap, Si$_3$N$_4$ waveguide circuit, for which two-photon absorption is negligible~\cite{moss_2013NatPhot}, coupled to an InP gain section to realize a hybrid integrated semiconductor laser with an intrinsic linewidth as low as 40~Hz. This is achieved by realizing a laser cavity of long photon lifetime, in spite of almost 100\% passive roundtrip loss, and in spite of high intracavity intensity. A scheme of the laser is displayed in Fig.~\ref{fig:hybrid_laser}, comprising an InP semiconductor amplifier and a dielectric, low-loss silicon nitride waveguide feedback circuit for cavity length extension. In this particular design, optical cavity length extension is obtained by a physical increase of length via a 33-mm long spiral in combination with an optical increase via resonant excitation of intracavity microring resonators. The end mirrors are the reflection at the back facet of the gain chip and the Sagnac loop mirror, meaning that the light passes the microring resonators twice per roundtrip. Narrow linewidth is achieved with three basic considerations. The first is providing a long photon lifetime already in a single roundtrip through a low-loss and long extension circuit. This decouples the laser cavity photon lifetime from intrinsically high loss in the remaining parts of the cavity, specifically, in the semiconductor amplifier, but also from loss resulting from coupling between different waveguide platforms, and due to strong output coupling for increased power efficiency.  Second, we exploit low propagation loss in the cavity extension to implement single-mode resolved spectral filtering already in a single roundtrip through the extension. This imposes single-mode oscillation with high side-mode suppression, which enables adjusting for stable laser operation at lowest linewidth without spectral mode hops. Third, to prevent that nonlinear loss does not compromise the photon lifetime, we use a wide-bandgap dielectric waveguide platform for laser cavity extension and restrict high-finesse spectral filtering solely to the dielectric part of the cavity. Thereby the laser linewidth can be decreased inversely with increasing the laser output.

\section{Conditions for narrow linewidth}
\label{sec:conditions}
To illustrate the key ingredients in our approach we recall the main conditions to induce narrow linewidth in extended cavity single-mode diode lasers~\cite{patzak_1983EL, henry_1986JLT, ujihara_1984JQE, bjork_1987JQE, koch_1990JLT}. The first condition is a long photon lifetime of the passive cavity because this increases the phase memory time of the laser resonator. If the total roundtrip loss can be reduced to below a few percent, the photon lifetime can be extended via multiple roundtrips in a short resonator~\cite{santis_2018PNAS}. In this case, due to the large free spectral range of short resonators, lower-finesse intracavity spectral filtering is sufficient for achieving single-mode oscillation. However, this approach is usually hard to realize due to intrinsically high passive roundtrip loss in semiconductor lasers.

\begin{figure}[tbp]
	\centering
		\includegraphics[width=0.6\linewidth]{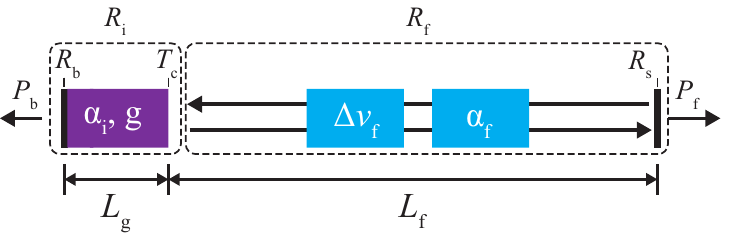}
	\caption{\label{fig:hybrid_laser_schematic} Schematic representation of a laser with an extended external cavity. The gain section has an intrinsic loss, $\alpha_{\textrm{i}}$, and gain, $g$, per unit length. Further, $L_{\textrm{g}}$ is the length of the gain section, $R_{\textrm{b}}$ is the back reflectivity and $T_{\textrm{c}}$ is the mode coupling efficiency at the interface. The passive gain section provides a total effective roundtrip reflectivity $R_{\textrm{i}}$. The feedback chip provides a low propagation loss, $\alpha_{\textrm{f}}$ ($\ll \alpha_{\textrm{i}}$), a long effective length, $L_{\textrm{f}}$, an end mirror with reflectivity $R_{\textrm{s}}$ and a spectral filtering with width $\Delta\nu_{\textrm{f}}$ to ensure single-mode oscillation, that are all combined in a single reflectivity $R_{\textrm{f}}=|r_{\textrm{f}}(\nu)|^2$, with $r_{\textrm{f}}(\nu)$ the total complex amplitude reflectivity of the feedback circuit. The large $L_{\textrm{f}}$ dominates the total length of the laser cavity and is responsible for increasing the photon lifetime and narrowing the laser linewidth, even in the presence of high intrinsic loss $\alpha_{\textrm{i}}$ (\textit{i.e.}, low $R_{\textrm{i}}$).} 
\end{figure}

\begin{figure}[tbp]
	\centering
	\includegraphics[width=0.6\linewidth]{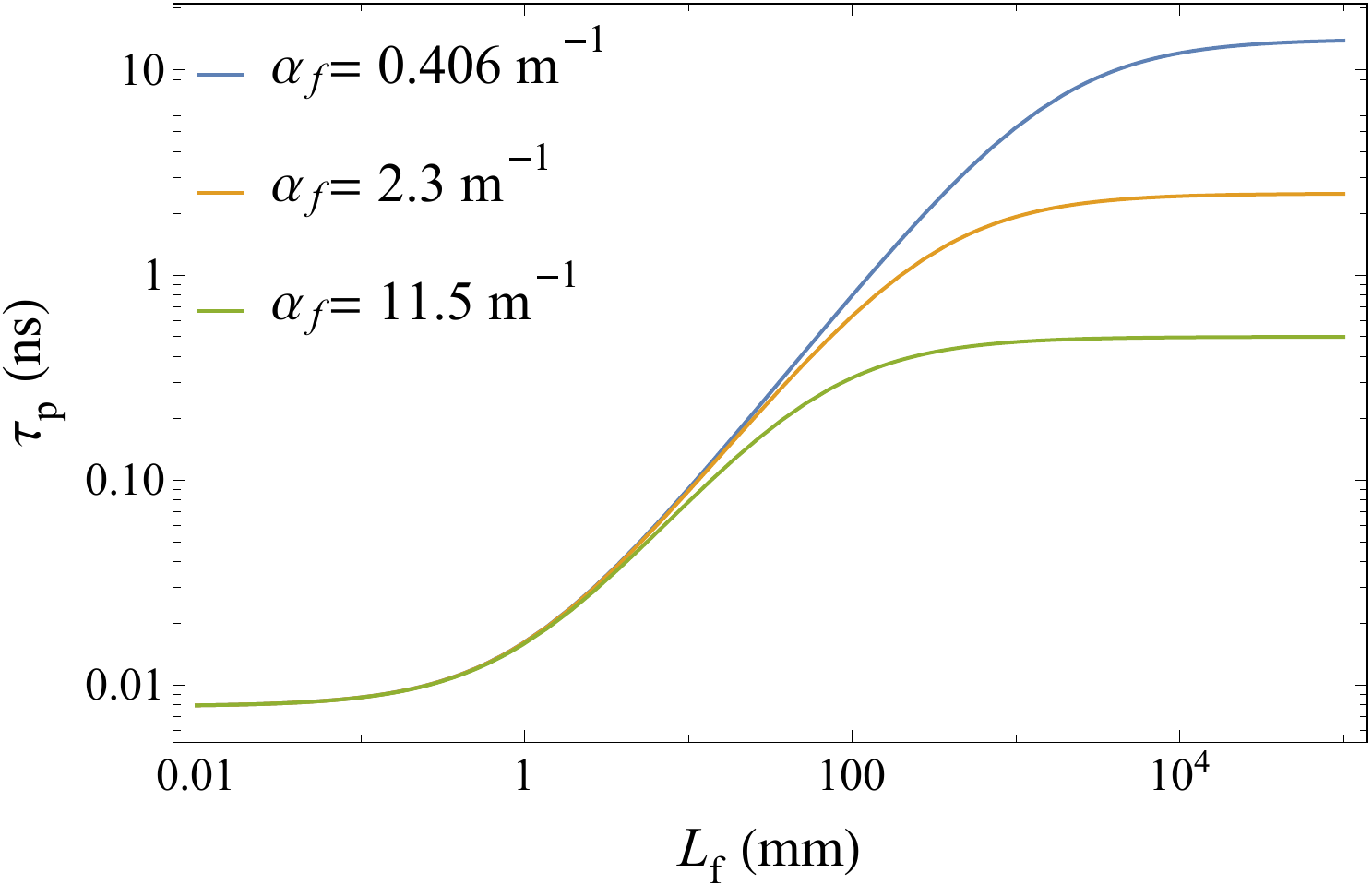}
	\caption{\label{fig:photon_lifetime}Calculated photon lifetime $\tau_{\textrm{p}}$ as a function of the geometric length $L_{\textrm{f}}$ of the feedback arm for a typical propagation loss of $\alpha_{\textrm{f}}=2.3$ m$^{-1}$ (0.1~dB/cm) and for losses that are a factor of 5 smaller and larger, while $\alpha_{\textrm{g}}=1600$~m$^{-1}$. Other parameters are $R_{\textrm{b}}=0.9$, $R_{\textrm{s}}=0.8$, $n_\textrm{g,i}=3.6$, $n_\textrm{g,f}=1.715$, and $L_{\textrm{g}}=1$ mm.}
\end{figure}

Our approach provides a long photon lifetime in spite of high passive roundtrip loss, by extending the laser cavity with a long feedback arm as displayed in Fig.~\ref{fig:hybrid_laser_schematic}. The laser comprises a gain section of length $L_{\textrm{g}}$ with intrinsic loss $\alpha_{\textrm{i}}$ and gain $g$ per unit length and a feedback arm of length $L_{\textrm{f}}$ having a propagation loss $\alpha_{\textrm{f}}$ per unit length. We define $L^{\textrm{(o)}}= n_{\textrm{g}} L$ as the group-index weighted optical length corresponding to a waveguide of length $L$ with effective group index $n_{\textrm{g}}$. The feedback arm also contains a narrow spectral filter with bandwidth $\Delta\nu_\textrm{f}$ to enable singe-mode oscillation. The end mirrors of the cavity have reflectances $R_{\textrm{b}}$ and $R_{\textrm{s}}$ through which a power $P_b$ and $P_f$ is extracted, respectively, from the laser cavity. The mode coupling at the interface between the gain section and feedback arm results in a transmittance $T_{\textrm{c}}$. To illustrate how the photon lifetime of the passive laser cavity changes with the length of the feedback arm, we assume, for simplicity, that $T_{\textrm{c}}=1$, \textit{i.e.}, we assume perfect coupling between the two sections, and that all microrings, that constitute the narrow band optical filter, are tuned to be perfectly resonant at the laser wavelength. Under these conditions, the photon lifetime $\tau_p$ is given by~\cite{coldren_2012}
\begin{equation}
    \frac{1}{\tau_p}= \frac{1}{L_{\textrm{g}}+L_{\textrm{f}}}\left(\alpha_{\textrm{i}}v_{\textrm{g,i}} L_{\textrm{g}} +\alpha_{\textrm{f}}v_{\textrm{g,f}} L_{\textrm{f}} - \frac{1}{2}\left\langle v_{\textrm{g}} \right\rangle\textrm{ln}(R_{\textrm{b}} R_{\textrm{s}})\right),
    \label{eq:photon_lifetime}
\end{equation}
where $v_\textrm{g,i}=c/n_\textrm{g,i}$ and $v_\textrm{g,f}=c/n_\textrm{g,f}$ are the effective group velocities of the gain and feedback section, respectively, with $c$ the speed of light in vacuum, $\alpha_m= -\textrm{ln}(R_{\textrm{b}} R_{\textrm{s}})/2(L_{\textrm{g}}+L_{\textrm{f}})$ is the distributed mirror loss and $\left\langle v_{\textrm{g}} \right\rangle = (v_{\textrm{g,i}} L_{\textrm{g}} +v_{\textrm{g,f}} L_{\textrm{f}})/(L_{\textrm{g}} +L_{\textrm{f}})$ is the length weighted average group velocity of the propagating optical mode. Taking as typical values $R_{\textrm{b}}=0.9$, $R_{\textrm{s}}=0.8$,  $\alpha_{\textrm{i}}=1600$~m$^{-1}$, $n_\textrm{g,i}=3.6$, $L_{\textrm{g}}=1$ mm and $n_\textrm{g,f}=1.715$, Fig.~\ref{fig:photon_lifetime} shows the calculated photon lifetime versus $L_{\textrm{f}}$ for a nominal propagation loss of $\alpha_{\textrm{f}}=2.3$ m$^{-1}$ (0.1~dB/cm) and for losses that are a factor of 5 smaller and larger, for comparison. Figure~\ref{fig:photon_lifetime} shows that for very small extension of the laser cavity, $L_{\textrm{f}} \ll L_{\textrm{g}}$, the photon lifetime is a constant. In this regime,  $1/\tau_p\approx v_{\textrm{g,i}}\left(\alpha_{\textrm{i}} - \frac{1}{2L_{\textrm{g}}}\textrm{ln}(R_{\textrm{b}} R_{\textrm{s}})\right)$, which is a constant independent of $L_{\textrm{f}}$. As $L_{\textrm{f}}$ increases, the photon lifetime starts to increase linearly with $L_{\textrm{f}}$ independent of the propagation loss $\alpha_{\textrm{f}}$ of the waveguide. In this regime $1/\tau_p\approx (1/L_{\textrm{f}})\left(\alpha_{\textrm{i}}v_{\textrm{g,i}} L_{\textrm{g}} - \frac{1}{2}v_{\textrm{g,f}}\textrm{ln}(R_{\textrm{b}} R_{\textrm{s}})\right) $ as $L_{\textrm{f}} \gg L_{\textrm{g}}$ and still $\alpha_{\textrm{f}} v_{\textrm{g,f}} L_{\textrm{f}} \ll \alpha_{\textrm{i}} v_{\textrm{g,i}} L_{\textrm{g}} - \frac{1}{2}v_{\textrm{g,f}}\textrm{ln}(R_{\textrm{b}} R_{\textrm{s}})$. Indeed, the photon lifetime is independent of $\alpha_{\textrm{f}}$ in this regime (see Fig.~\ref{fig:photon_lifetime}). This means that in stationary state the gain coefficient of the hybrid laser only weakly depends on the propagation loss of the feedback section and that the amount of spontaneous emission, which is the source for the intrinsic linewidth, is approximately constant when increasing $L_{\textrm{f}}$. The resulting increase in phase memory time corresponds to a narrowing of the intrinsic linewidth.  Furthermore, if we define $R_\textrm{i}=R_{\textrm{b}} e^{-2\alpha_{\textrm{i}} L_{\textrm{g}}} T_{\textrm{c}}^2$ as the reflectance of the passive gain section and $R_{\textrm{f}}(\nu) = |r_{\textrm{f}}(\nu)e^{i\phi_{\textrm{f}}(\nu)}|^2$, with $r_{\textrm{f}}(\nu)e^{i\phi_{\textrm{f}}(\nu)}$ the frequency dependent effective complex amplitude reflectivity of the feedback arm (see Fig.~\ref{fig:hybrid_laser_schematic}), we have $R_{\textrm{f}} \gg R_{\textrm{i}}$ for typical values of $R_{\textrm{s}}$ used in our experiment. We define this as the strong feedback regime. If $L_{\textrm{f}}$ is still further increased, the total propagation loss will eventually become the dominant loss leading to a saturation of the photon lifetime. In this regime $1/\tau_p \approx v_{\textrm{g,f}}\alpha_{\textrm{f}}$. This is clearly visible in Fig.~\ref{fig:photon_lifetime} for the different propagation losses. For the nominal propagation loss, the photon lifetime saturates at about 2.5~ns for $L_{\textrm{f}} \gtrsim 1$~m. Note that we have kept the mirror reflectances constant and only changed the effective  length of the feedback arm.

Including a more realistic $T_{\textrm{c}}$ does not change the above observations as a proper design of the hybrid laser will lead to $T_{\textrm{c}} \gtrsim 0.9$. Including the associated loss in the intrinsic loss $\alpha_{\textrm{i}}$ of the gain section only slightly increases this value. From this we conclude that extending the length of the feedback arm above a threshold, that, for a given outcouple loss, is set by the length of the gain section, will linearly increase the photon lifetime, and hence reduce the intrinsic linewidth, as long as the total loss, due to passive losses in the gain section and outcoupling, dominates the total propagation loss in the feedback circuit.

Including off-resonance effects of the filter on the photon lifetime, and thus on the intrinsic linewidth of the laser, is more complicated as the total feedback loss increases and becomes strongly frequency dependent, while at the same time the length of the feedback arm reduces when the filter is detuned from the line center. To illustrate the effect on the instrinsic linewidth, we consider the whole feedback arm as a lumped reflectance $R_{\textrm{f}}(\nu)$ (see Fig.~\ref{fig:hybrid_laser_schematic}) and recall the expression for the intrinsic or Schawlow-Townes linewidth $\Delta\nu_{\textrm{ST}}$~\cite{fan_2017OE}
\begin{equation} \label{eq:schawlow_townes}
    \Delta\nu_{\textrm{ST}}=\frac{h\nu}{4\pi}\frac{n_{\textrm{sp}} \gamma_{\textrm{tot}} \gamma_{\textrm{m}} F_{\textrm{P}}}{P_0 K(\nu)} \frac{1+\alpha_{\textrm{H}}^2}{F^2}.
\end{equation}
Here, $\gamma_{\textrm{m}}=-\frac{v_{\textrm{g}}}{2L_{\textrm{g}}} \textrm{ln}(R_{\textrm{b}} R_{\textrm{f}}(\nu))$ is the mirror loss rate and $\gamma_{\textrm{tot}}$ is the total loss rate, both assumed to be homogeneously distributed over the length of the gain section. Further, $h\nu$ is the photon energy of the laser oscillation, $P_0$ the output power at a particular output port, and $K(\nu)>1$ a weight factor accounting for power emitted from other ports. $F_{\textrm{P}} >1$ is the longitudinal Petermann factor increasing the linewidth, in case that reflective feedback ($R_{\textrm{b}}$ and $R_{\textrm{f}}(\nu)$) becomes very small~\cite{ujihara_1984JQE, henry_1986JLT}. $\alpha_{\textrm{H}}$ is the Henry linewidth enhancement factor due to gain-index coupling~\cite{henry_1982JQE} and $n_{\textrm{sp}}$ is the spontaneous emission enhancement factor that takes into account the reduction in inversion due to reabsorption by valence band electrons. Typically, $n_{\textrm{sp}}$ takes a value of around 2. Finally, $F=1+A+B$, where $A=\frac{1}{\tau_{\textrm{g}}}  \frac{d}{d\nu}\phi_{\textrm{f}}(\nu)$, $B=\frac{\alpha_{\textrm{H}}}{\tau_{\textrm{g}}}\frac{d}{d\nu} \textrm{ln}(|r_{\textrm{f}}(\nu)|)$ and $\tau_{\textrm{g}}$ is the roundtrip time of the solitary gain section. At resonance, \textit{i.e.}, when the center of the filter's reflection peak coincides with the oscillation frequency, $B=0$ and $A$ is  maximum and equal to the ratio of the optical length of the feedback arm to the optical length of the gain section~\cite{fan_2017OE, tran_2020JSTQE}. We find that $\Delta\nu_{\textrm{ST}}$ reduces with the inverse of $L_{\textrm{f}}^2$ when keeping the end mirror reflectances constant, in agreement with our discussion on the photon lifetime above. Off-resonance, $A$ decreases and $B$ increases on the rising edge of the filter peak whenever gain-index coupling is present, \textit{i.e.}, whenever $\alpha_{\textrm{H}}>0$. For a sufficiently sharp reflection peak, the maximum in $F$ is found for an oscillation frequency on the rising edge of the filter's reflection peak, slightly detuned from the line center, where spontaneous emission-induced index and frequency fluctuations are compensated by a steep, frequency dependent resonator loss~\cite{kazarinov_1987JQE, olsson_1987APL, koch_1990JLT, tran_2020JSTQE}. 

A full numerical analysis of Eq.~(\ref{eq:schawlow_townes}), including changes in $R_{\textrm{f}}(\nu)$ and, consequently, $K(\nu)$ and $F_{\textrm{P}}$ for a more complete off-resonance description of the linewidth behaviour, is still of limited value due to the underlying assumptions used in deriving Eq.~(\ref{eq:schawlow_townes}). For example, the spatial distribution of the inversion density varies notably along the axis of the gain section, which is due to the relatively high intrinsic loss $\alpha_{\textrm{i}}$. This means that the mean field approximation used in deriving Eq.~(\ref{eq:schawlow_townes}) is not well justified. While not suitable for accurate predictions of the intrinsic linewidth for our hybrid diode laser, Eqs.~(\ref{eq:photon_lifetime}) and (\ref{eq:schawlow_townes}) are still very useful to determine scaling properties and design strategies for lowering the intrinsic linewidth of the laser. We have shown that increasing the optical length of the feedback arm narrows the linewidth as long as the total propagation loss in the feedback arm is not the dominant loss in the laser, \textit{i.e.}, the total propagation loss, including nonlinear loss, is only a small fraction of the remaining loss. This means that the maximum obtainable photon lifetime is set by the propagation loss of the feedback arm.

The second condition is a sufficiently high spectral resolution of the feedback filter for single-mode oscillation to allow the use of  Eq.~(\ref{eq:schawlow_townes}). Furthermore, compensating the linewidth enhancement due to gain-index coupling via strongly frequency selective loss at the low-frequency side of the feedback filter transmission requires fine tuning of the laser without spectral mode hops. Such a fine-tuning requires single-mode resolved spectral filtering in the feedback arm. Single-mode filtering is obtained if the FWHM bandwidth of the reflection peak of the filter, $\Delta\nu_f$, is narrower than the laser cavity mode spacing.

The third condition for narrow linewidth is operating the laser maximally high above threshold. This reduces the relative rate of randomly phased spontaneous emission as compared to phase-preserving stimulated emission. At a given roundtrip loss, high-above-threshold operation can only be achieved by increasing the pump power. In the experiment, we increase the laser power for linewidth narrowing. To maintain single-mode oscillation with high-finesse spectral filtering, we use a dielectric waveguide platform for extending the cavity length, where spectral filtering is implemented only with dielectric materials. This choice ensures that high intracavity intensity, occurring at high laser power due to filter-induced enhancement, is only present in the dielectric part of the laser. There, nonlinear loss can be safely neglected~\cite{moss_2013NatPhot, kuyken_2017NANOPHOT} due to the wide bandgap of dielectric materials.

\section{Laser design}
Figure~\ref{fig:hybrid_laser} shows the schematic design of the hybrid laser, comprising an InP semiconductor optical amplifier (gain section) and an extended cavity made of a long Si$_3$N$_4$ low-loss dielectric waveguide circuit that provides frequency selective feedback to the amplifier. As the directional couplers and, to a lesser extent, the effective refractive index of the waveguide used in the laser design are wavelength dependent, in the following the nominal wavelength of 1560~nm is assumed unless otherwise specified.

The InP semiconductor amplifier (COVEGA, SAF 1126) for generation of light in a single transverse mode at around 1.55~$\mu$m wavelength has a length of $L_{\textrm{g}}=1000$~$\mu$m  and a specified typical output power of 60~mW based on amplification in multiple quantum wells. The back facet is high-reflection coated ($R_{\textrm{b}}=90\%$) to provide double-pass amplification. In order to suppress back-reflection into the amplifier, the front facet is anti-reflection coated to a specified reflectivity of $10^{-4}$ for an external index of 1.5, which is close to the effective refractive index of the tapered input end of the  Si$_3$N$_4$  waveguide circuit (1.584). The semiconductor waveguide is tilted by 6$^{\circ}$, to further reduce back-reflection. Derived from the far-field specifications, the mode field diameter at the exit facet is 4.4~$\mu$m in the horizontal and 1.3~$\mu$m in the vertical direction. The amplifier is integrated with the Si$_3$N$_4$ circuit via alignment for maximizing the amount of amplified spontaneous emission (ASE) entering the Si$_3$N$_4$ circuit, followed by bonding with an adhesive. The integrated laser is mounted on a thermoelectric cooler and kept at 25~$^{\circ}$C. The electrical connects are wire bonded to a fan-out electronic circuit board. For driving the amplifier with a low-noise current, we use a battery-driven power supply (ILX Lightwave, LDX3620).
\begin{figure}[tbp]
	\centering
		\includegraphics[width=0.4\linewidth]{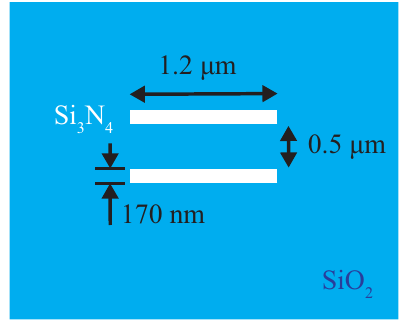}
	\caption{\label{fig:double_stripe} Schematic view of the cross section of the double stripe Si$_{3}$N$_{4}$ waveguide used in the photonic feedback circuit for the hybrid laser. The supported single transverse optical mode has a cross section of $1.6 \times 1.7$~$\mu$m$^2$.} 
\end{figure}

A long optical path length for linewidth narrowing, and sharp spectral filtering for single-mode oscillation, is provided with a Si$_3$N$_4$ circuit optimized for low-loss and high frequency selectivity. In this platform~\cite{roeloffzen_2018JSTQE} the core cross section can be adjusted to obtain a proper combination of tight guiding and low loss. We select a symmetric double-stripe geometry, see Fig.~\ref{fig:double_stripe}, that comprises two Si$_3$N$_4$ cores (1.2~$\mu$m $\times$ 170~nm) separated by 500~nm embedded in a SiO$_2$ cladding. This cross section yields a single-spatial mode of size $1.6 \times 1.7$~$\mu$m$^2$ for the TE polarization and an effective group index of 1.715. The propagation loss is smaller than 0.1~dB/cm, which agrees with values reported by Roeloffzen \textit{et al}.~\cite{roeloffzen_2018JSTQE}, and is determined from light scattering measurements with an IR camera using test structures from the same wafer with lengths of 5, 10 and 15~cm. The chosen cross section and the high index contrast between core and cladding ($\Delta n \approx 0.53$) provides tight guiding, making radiative loss (bending loss) negligible also for waveguides with tight bending radii as small as 100~$\mu$m. This enables to employ small-radius, low-loss microring resonators for Vernier-filtering with a wide free spectral range (FSR) comparable to the gain bandwidth~\cite{oldenbeuving_2013LPL}. Tight guiding in combination with low loss enables to realize significant on-chip optical path lengths. For example, extending the cavity length such that the returning power drops to a fraction of $R_{\textrm{f}}=1/3$ and assuming nominal parameters otherwise, \textit{i.e.}, a Sagnac mirror reflectance of $R_{\textrm{s}}=0.9$, a loss coefficient of $\alpha_{\textrm{f}}=0.1$~dB/cm, and a ring power coupling of $\kappa^2=10\%$, results in a roundtrip optical length of the laser cavity of about $2L^{\textrm{(o)}}=74$~cm. This corresponds to extending the photon lifetime to about 1 nanoseconds (see Fig.~\ref{fig:photon_lifetime}). The selected waveguide cross section is also suitable for low-loss adiabatic tapering. With two-dimensional tapering, the calculated maximum power coupling to the mode field of the gain section is in the range of $T_{\textrm{c}}$=90 to 93\% ~\cite{fan_2016PJ}, and the coupling to the $10.5 \pm 0.8 \mu$m diameter mode of single-mode output fibers (Fujikura 1550PM) can be as high as 98\%. 

\begin{figure}[tbp]
	\centering
		\includegraphics[width=0.9\linewidth]{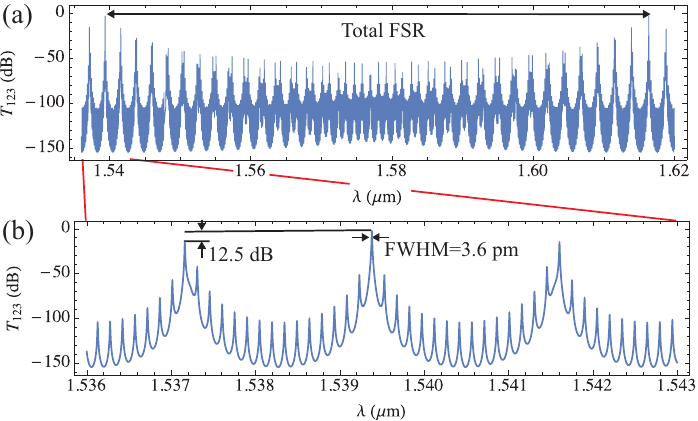}
	\caption{\label{fig:transmission} Calculated double-pass power transmission $T_{123}$ of the Si$_3$N$_4$ feedback arm containing three cascaded rings with radii $R_1=99$~$\mu$m, $R_2=102$~$\mu$m and $R_3=1485$~$\mu$m across a range corresponding to the gain bandwidth (a) and across a small range near the maximum of the gain at 1.54~$\mu$m  (b). The peak transmission amounts to 51\% as calculated with an effective group index of $n_g=1.715$, the Sagnac mirror reflectance set to 90\%, and a propagation loss of $0.1$~dB/cm.} 
\end{figure}

At this point we recall that we do not aim on low loss per entire roundtrip through the hybrid cavity. Instead we maximize only the optical length and thus the photon travel time in the dielectric feedback arm of the laser cavity, while keeping the loss in the feedback arm much lower than the intrinsic loss in the remaining part of the laser cavity roundtrip. With the circuit design realized here, the feedback arm provides a high peak reflectivity of $R_{\textrm{f}}=51\%$, assuming the Sagnac mirror reflectance is set to $R_{\textrm{s}}=90\%$ and using nominal values for the propagation loss of $\alpha_{\textrm{f}}=0.1$~dB/cm and power coupling of the rings of $\kappa^2=10\%$. In contrast, the loss in the remaining parts of the laser cavity is much higher, \textit{i.e.}, $R_{\textrm{i}} \approx 3\%$. The latter is calculated from double-passing 80\% loss in the amplifier,  10\% loss at at the amplifier back facet,  and double-passing 10\% loss at the InP-Si$_3$N$_4$ interface. The loss estimates show that the laser would operate in the strong feedback regime, where $R_{\textrm{f}} \gg R_{\textrm{i}}$, such that a long roundtrip length in the feedback circuit should enable significant linewidth narrowing. 

In order to induce single frequency oscillation across the 70~nm (8~THz) wide gain bandwidth in spite of an expected, dense mode spacing of a long laser cavity, we use three cascaded microring resonators in add-drop configuration, all with a power coupling of $\kappa^2=10\%$ to their bus waveguides.  Two short resonators with a small difference in radius are used in Vernier configuration for coarse frequency selection ($R = 99$ and 102~$\mu$m, average FSR 278~GHz, finesse~28, quality factor $Q \approx 20,000$). The third microring resonator provides fine spectral filtering ($R_3=1485$~$\mu$m, FSR 18.6~GHz, finesse~28, $Q \approx 290,000$). Taking into account that all resonators are double-passed in the silicon nitride feedback circuit and assuming 0.1~dB/cm propagation loss, we calculate a FWHM of the spectral filter's transmission peak of 450~MHz (3.6~pm). Behind the resonators the extended cavity is closed with a Sagnac loop mirror of adjustable reflectivity via a tunable balanced Mach-Zehnder interferometer.
For a fixed setting of the mirror, the fraction of power coupled out of the laser cavity typically varies less than 20\% over the gain bandwidth of the laser. However, by tuning the Mach-Zehnder interferometer, the outcoupling can approximately be kept constant in the experiment when the oscillating wavelength is varied. The output power is collected into a single-mode output fiber (Fujikura 1550PM) that is butt-coupled with index matching glue to the polished end facet of the feedback chip.

For spectrally aligned and resonant microring resonators, we calculate a laser cavity optical roundtrip length of $2L^{\textrm{(o)}} = 0.49$~m which, via $\textrm{FSR}=c/2L^{\textrm{(o)}}$, $c$ being the speed of light in vacuum, corresponds to a free spectral range of 610~MHz. The length is calculated with double-passing the optical length of the three resonators, each having a power-coupling of $\kappa^2=10\%$ (which corresponds to multiplying each length with the approximate number of nine round trips at resonance \cite{liu_2001APL}), a 33~mm long waveguide spiral for further cavity length extension, the length of the amplifier, and various smaller sections of bus waveguides including the loop mirror (all geometric lengths are converted to optical lengths). With this cavity length the passive cavity photon lifetime already starts to saturate (see Fig.~\ref{fig:photon_lifetime}). We  note that the cavity mode spacing varies noticeably with the light frequency, which is mainly due to strong dispersion in transmission through the long microring resonator. For light at transmission resonance of the long resonator, this places the two closest cavity modes at 965~MHz distance. For light in the midpoint wing of the transmission resonance, the closest cavity mode is located at 750~MHz. In comparison, the 450-MHz bandwidth of the feedback filtering is smaller, \textit{i.e.}, the condition of single-mode resolved filtering is fulfilled. We further note that, based on measurements on similar structures, the power coupling $\kappa^2$ of the 10\% directional coupler typically increases by almost a factor of 2 when the wavelength increases from 1500 to 1600~nm. This means that the cavity length extension is longer at the short wavelength side of the gain bandwidth. As for these lengths the cavity photon lifetime already starts to saturate, we expect a reduced variation in cavity photon lifetime over this gain bandwidth. 

The calculated double-pass filter spectrum obtained with the three-ring circuit across a range corresponding to the gain bandwidth is shown in Fig.~\ref{fig:transmission}(a) and across a small range around the resonant wavelength of 1.5359~$\mu$m in Fig.~\ref{fig:transmission}(b). For a Sagnac mirror reflectivity of 90\%, as used for spectral recordings, we calculate a high feedback of $R_{\textrm{f}}=51\%$, which is due to low loss in the Si$_3$N$_4$ waveguides. The feedback at the next-highest side resonance of the long resonator is lower by -12.5 dB. For setting the highest transmission peak to any laser cavity mode within the laser gain, the resonators are equipped with thin-film thermo-electric phase shifters with a $0-2\pi$ range. With the described spectral filtering and due to the dominance of homogeneous gain broadening in the quantum well amplifier, it is expected that single-mode oscillation with high side mode suppression ratio is possible at any wavelength within the gain bandwidth. 

\section{Results}
\begin{figure}[btp]
	\centering
		\includegraphics[width=0.85\linewidth]{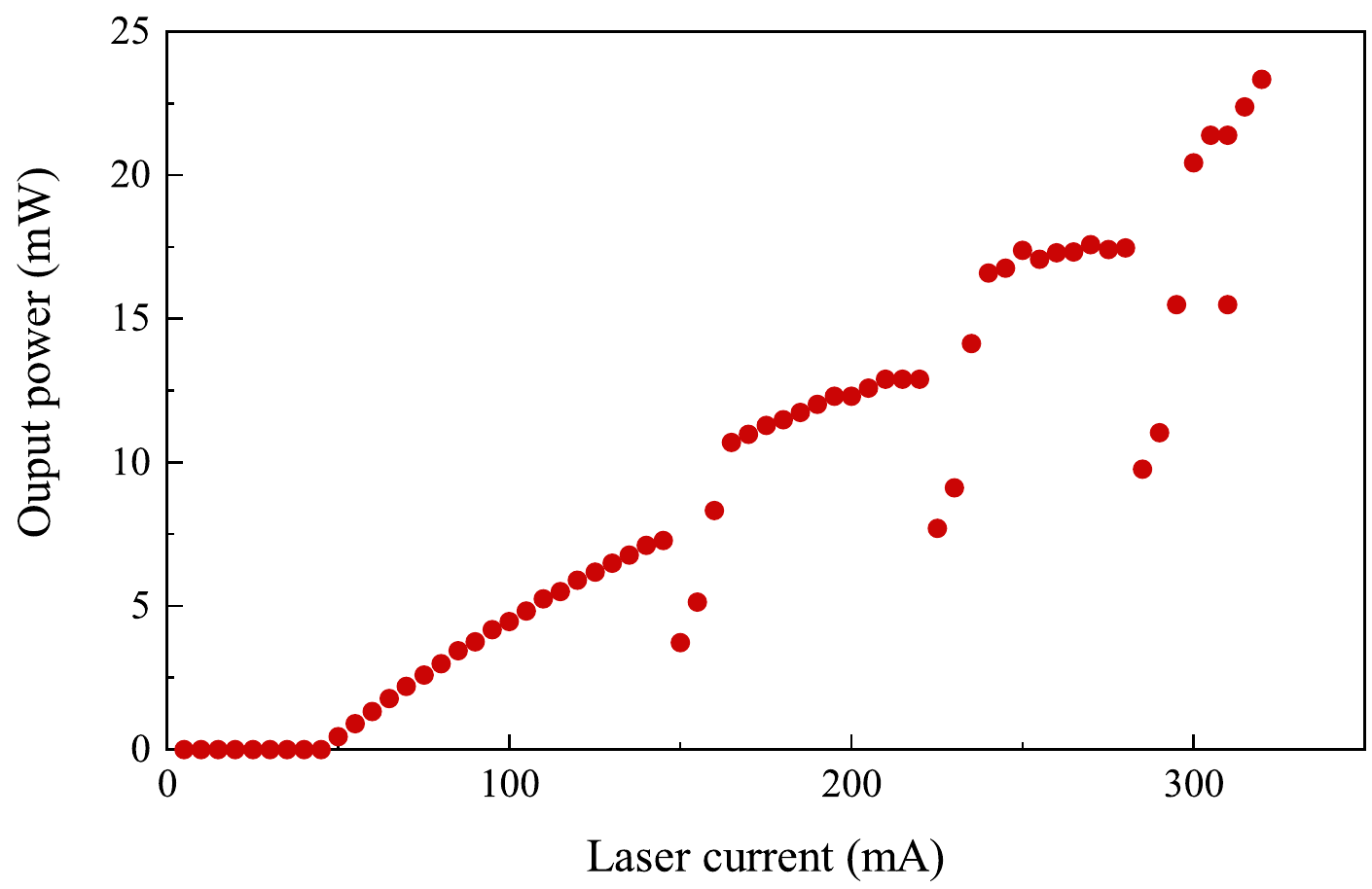}
	\caption{\label{fig:PI_curve} Typical laser output power as measured with increasing pump current, yielding a maximum output of 23 mW. The discontinuities indicate spectral mode hops. This particular measurement was performed at a wavelength of 1561~nm.} 
\end{figure}
Figure~\ref{fig:PI_curve} shows a typical measurement of the fiber-coupled output power behind the Sagnac loop mirror versus pump current. For achieving high output, the Sagnac mirror was set to a high transmission of about 80\%, and the laser wavelength was set to 1561~nm, near the center of the gain spectrum, via the phase shifter of the first microring resonator. The pump current is stepwise varied and fine-tuned, in order to maintain single-mode operation. The laser shows a threshold pump current of about 42~mA and a maximum output power of 23~mW is achieved at a pump current of 320~mA. This is almost half of the specified maximum power of the amplifier of 60~mW. The discontinuities in the output power versus pump current correspond to spectral mode hops. The reason is that increasing the pump current also changes the refractive index in the amplifier, which tunes the laser cavity length with regard to the transmission spectrum of the feedback filter.

To discuss the presence of nonlinear loss, we estimate the maximum intracavity intensity that occurs at the maximum output power. Assuming a Sagnac mirror transmission of 10\%, which is typically used for the linewidth measurements, we calculate a power of approximately 4~W in the largest microring resonator (2~W in each direction). Using a mode area of $1.6 \times 1.7$~$\mu$m$^2$ the according intensity is high, of the order of 0.15~GW/cm$^2$. However, loss from two-photon absorption can safely be neglected~\cite{moss_2013NatPhot} due to the wide bandgap of Si$_3$N$_4$. For comparison, in a silicon waveguide the same power and a typical mode field area of $0.5 \times 0.5$~$\mu$m$^2$ would cause significant two-photon absorption, \textit{i.e.}, of the order of 5~dB/cm~\cite{kuyken_2017NANOPHOT}. This would make it difficult to implement sharp spectral filtering to realize long, resonator-enhanced, feedback lengths and to narrow the intrinsic linewidth via increasing  the laser power.


\begin{figure}[tbp]
	\centering
		\includegraphics[width=0.85\linewidth]{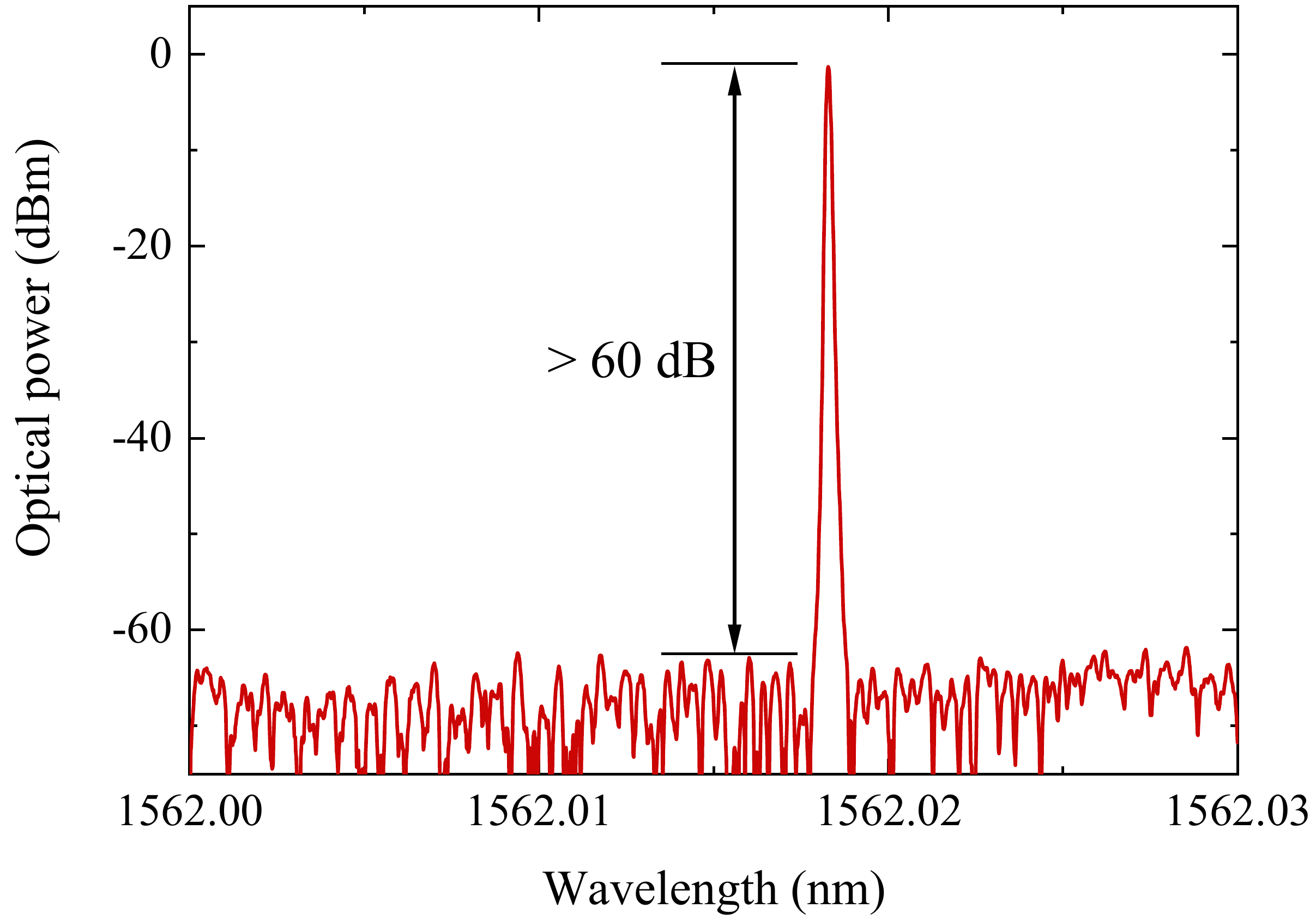}
	\caption{\label{fig:single_line_BOSA} Typical power spectrum recorded across a range of 30~pm with 0.1~pm resolution (3.7~GHz and 12~MHz, respectively).} 
\end{figure}
To verify that the laser oscillates at a single wavelength, the laser output spectrum is measured at the fiber-coupled output from the through port of the first small resonator (monitor port in Fig.~\ref{fig:hybrid_laser}). There it would be possible to observe also light that is not resonant with the microring resonators. To obtain a higher resolution than the mode spacing, the laser spectrum was recorded with an optical spectrum analyzer based on stimulated Brillouin scattering (Aragon Photonics, BOSA400), and the small resonators are tuned for single-mode oscillation. All spectral measurements are performed behind an optical isolator and using tilted fiber connections to avoid feedback into the laser. Figure~\ref{fig:single_line_BOSA} shows a typical power spectrum recorded with the maximum resolution of 0.1~pm (12~MHz) across a 30~pm (3.7~GHz) wide interval around the oscillating mode. This range spans four to five mode spacings, such that possibly oscillating side modes would have become detectable. The measured spectrum confirms clean single-mode oscillation, with a side mode suppression of about 62~dB. Using a second optical spectrum analyzer (ANDO, AQ6317), set at a lower resolution of 50~pm but larger scan range, we confirmed single mode oscillation over the complete gain bandwidth.
\begin{figure}[tbp]
	\centering
		\includegraphics[width=0.85\linewidth]{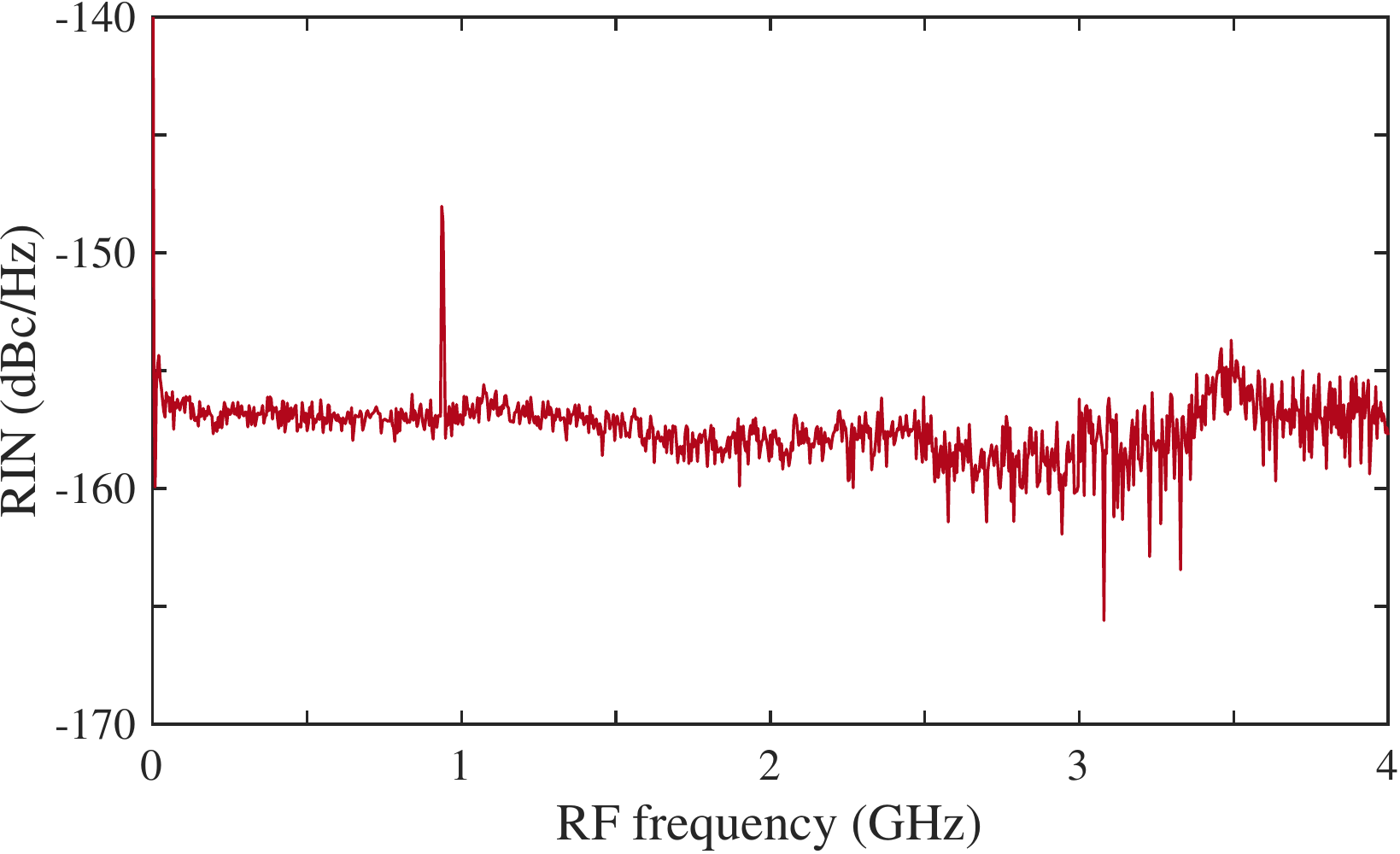}
	\caption{\label{fig:RIN} Typical power spectrum of the relative intensity noise (RIN). The spectrum is flat except for a small intermittent peak around 950~MHz. The optical output power was 1.2~mW.} 
\end{figure}

For further characterization we measure relative intensity noise (RIN) with a fast photodiode and RF spectrum analyzer (10~kHz resolution and 100~kHz video bandwidth). Figure~\ref{fig:RIN} shows a typical RIN spectrum when the optical output power was 1.2~mW, which displays flat, broadband and low intensity noise around -157~dBc/Hz. Single narrowband features, here at 940 MHz, are likely due to spurious RF pickup, as not all spectra display these.   
\begin{figure}[tbp]
	\centering
		\includegraphics[width=0.85\linewidth]{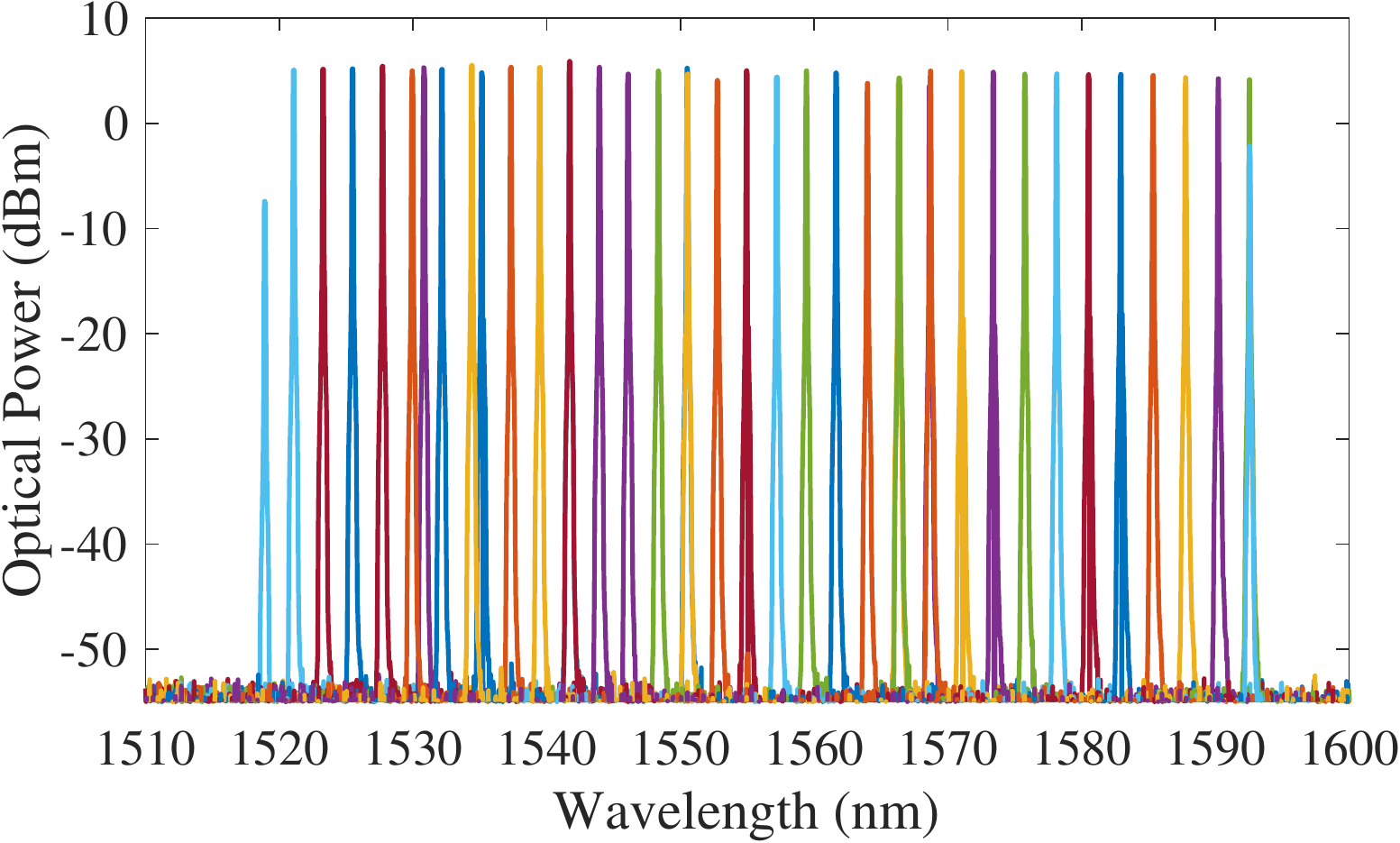}
	\caption{\label{fig:tuning_coarse} Superimposed output spectra recorded by tuning the laser wavelength in steps of 2~nm across a range of $>70$~nm.} 
\end{figure}

\begin{figure}[tbp]
	\centering
		\includegraphics[width=0.85\linewidth]{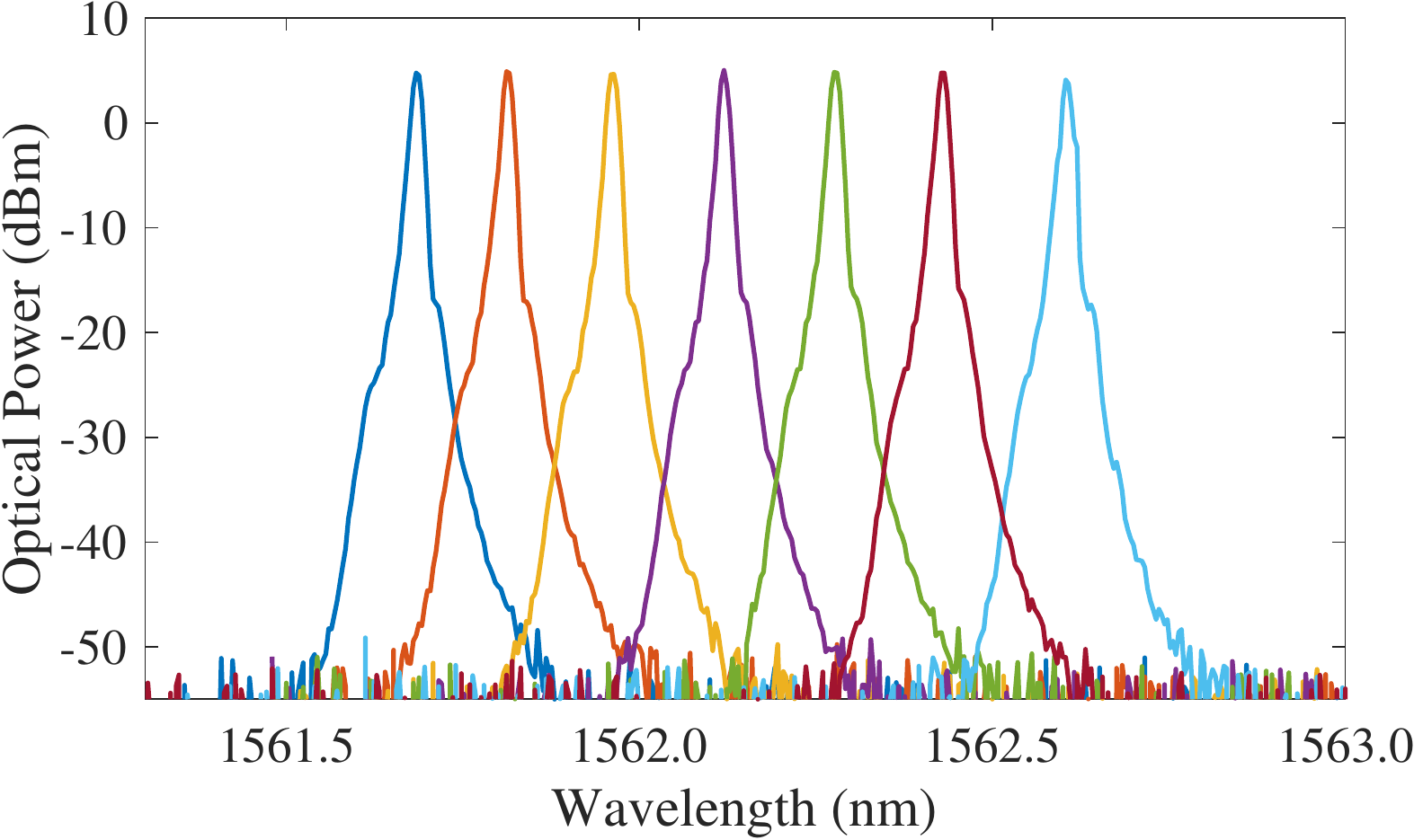}
	\caption{\label{fig:tuning_fine} Superimposed spectra when fine tuning the laser in steps of 0.15~nm.} 
\end{figure}

To explore the overall spectral coverage of single-mode oscillation, the laser was manually tuned via the phase shifters on top of the microring resonators using a maximum heater power of 270~mW per heater. Figure~\ref{fig:tuning_coarse} shows an example of superimposed laser spectra, with the laser tuned to 35 different wavelengths. For coarse wavelength tuning, the heater current of one of the small microresonators is increased. This gives rise to discrete wavelength changes at a stepsize of about 2~nm, which corresponds to the FSR of the other small resonator. After the wavelength is set to a desired value, also the heating current of the other small resonator is adjusted for maximum laser output, to improve the spectral alignment of all resonators. The approximately flat tuning envelope is obtained by adjusting the Sagnac mirror feedback with wavelength tuning, at a pump current of 200~mA. We obtain a spectral coverage of 74~nm and at least 3 mW of output power. This compares well with the current record for monolithic, heterogeneously and hybrid integrated lasers~\cite{latkowski_2015PJ,fan_2017CLEO, tran_2020JSTQE}. Fine-tuning shown in steps of the FSR of the large microring resonator is shown in Fig.~\ref{fig:tuning_fine}. This was achieved via tuning the small resonators and loop mirror without heating the long resonator.

The intrinsic linewidth of the laser is measured using two independent setups based on delayed self-heterodyne detection ~\cite{richter_1986JQE,mercer_1991JLT}. The first, a proprietary setup, uses a Mach-Zehnder interferometer with 5.4~m optical arm length difference, a 40-MHz acousto-optic modulator, and two photodiodes for balanced detection. The beat signal is recorded versus time and analyzed with a computer to obtain the power spectral density of frequency noise (PSD). Free-running lasers, as investigated here, typically display increased technical noise at low frequencies whereas, at high noise frequencies, the PSD level levels off to the intrinsic laser linewidth. The second uses an arm length difference of 20~km and an 80-MHz modulator (AA  Opto  Electronic, MT80-IIR60-F10-PM0.5-130.s with ISOMET 232A-2 AOM driver). The power spectrum of the beat signal is recorded with an RF spectrum analyzer (Agilent E4405B with 25~kHz RBW). The intrinsic linewidth is retrieved with Lorentzian fits to the line wings where the Lorentzian shape is minimally obstructed, \textit{i.e.}, avoiding the low-frequency noise regime near the line center, as well as the range close to the electronic noise floor. Linewidth measurements are carried out at various different pump currents at a wavelength of 1561~nm near the center of the gain spectrum. 

\begin{figure}[tbp]
	\centering
		\includegraphics[width=0.85\linewidth]{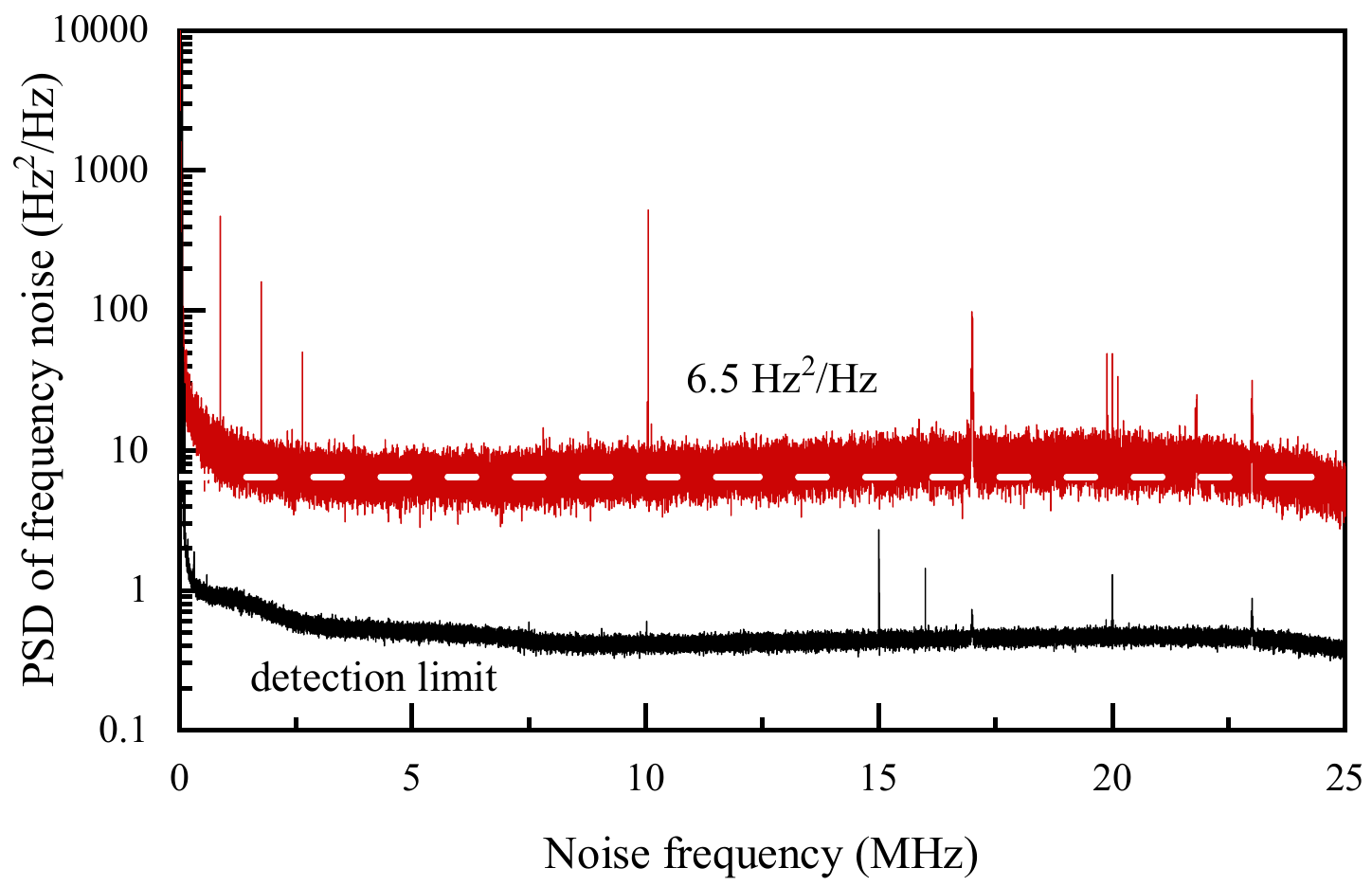}
	\caption{\label{fig:noise_PSD}Double-sided power spectral density (PSD) of laser frequency noise for a pump current of 255~mA, plotted for positive frequencies. The dashed line at 6.5 Hz$^2$/Hz represents the mean of PSD values for noise frequencies between 4 and 7.5~MHz. The detection limit is at 0.5~Hz$^{2}$/Hz.} 
\end{figure}

Figure~\ref{fig:noise_PSD} shows the PSD measured at a pump current of 255~mA, after adjusting for lowest noise only via the small microring resonators, while also monitoring the optical spectrum with an OSA to verify single-mode oscillation. The laser noise spectrum becomes flat for noise frequencies above $>2$~MHz. The upper bound for the white noise limit, indicated as dashed line, is taken as $6.5 \pm 1.3$~Hz$^2$/Hz. These values are obtained by taking the mean value and standard deviation of the Gaussian distribution of PSD values between noise frequencies of 4 and 7.5 MHz. After multiplying with $2\pi$ this corresponds to an intrinsic linewidth of $40 \pm 8$~Hz. This is significantly lower than our previous result of 290~Hz~\cite{fan_2017CLEO}. The lower linewidth has been obtained by using a different gain section, the COVEGA SAF 1126 InP gain chip, and by using a different outcoupling by the Sagnac loop mirror, about 10\% instead of 50\%, resulting in a doubling of the fiber coupled output power.

\begin{figure}[tbp]
	\centering
		\includegraphics[width=0.85\linewidth]{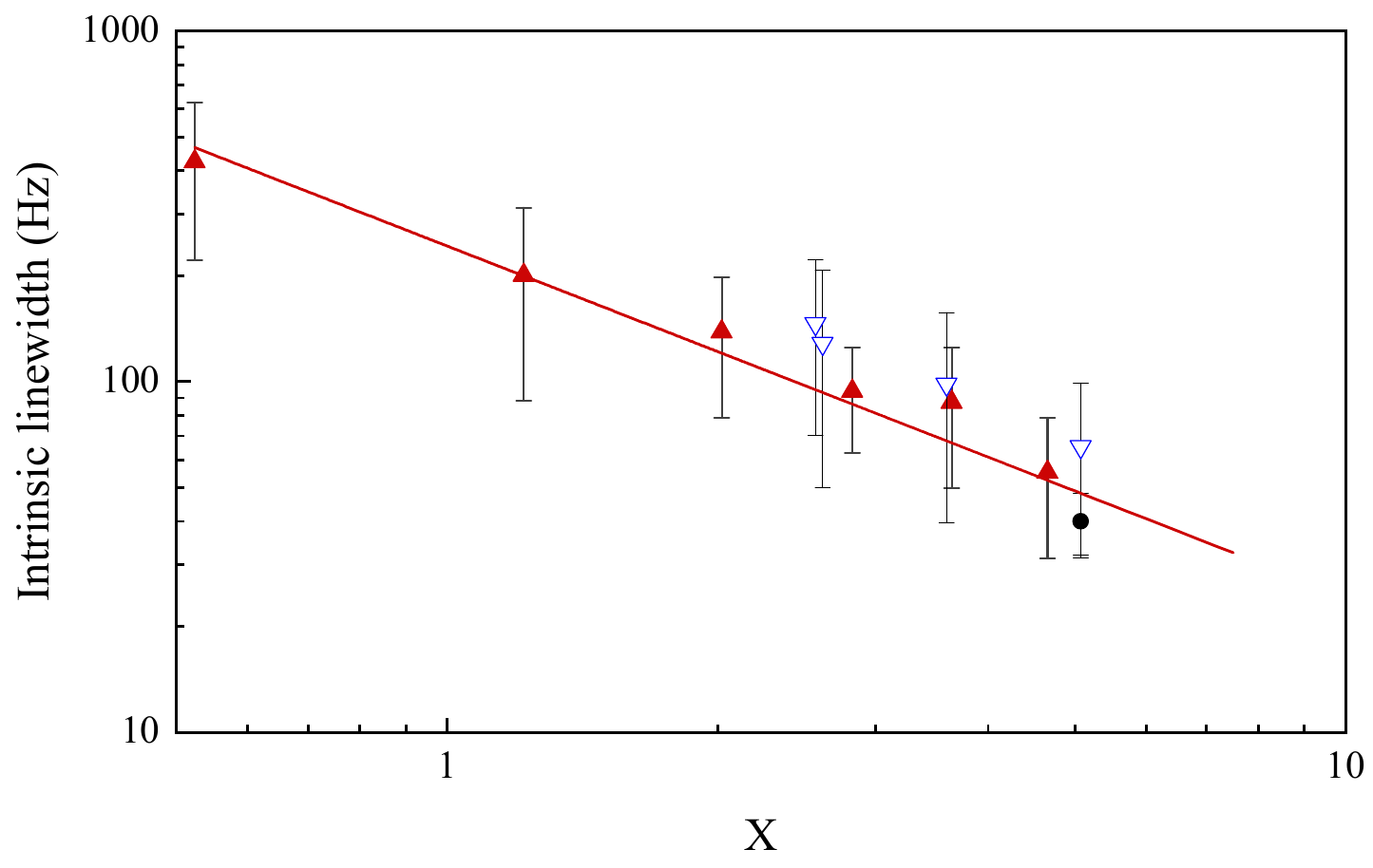}
	\caption{\label{fig:lorentzian_width}Double logarithmic plot of Lorentzian linewidth versus the threshold factor, $X=(I_{\textrm{p}}-I_{\textrm{p,th}})/I_{\textrm{p,th}}$, which is proportional to the output power, $P_{\textrm{out}}$. Unfilled symbols show measurements vs decreasing power. Measurements vs increasing power (filled symbols) yield slightly smaller linewidths. The solid line is a least-square fit to the lower linewidth data with negative unity slope (inverse power law, $\propto P_{\textrm{out}}^{-1}$). The linewidth obtained from PSD measurements (Fig.~\ref{fig:noise_PSD}) is shown as a black round symbol at $X=5.07$ (255~mA pump current).} 
\end{figure}

To verify the low linewidth level, the measurement is repeated with the second heterodyne setup using the same heater settings. The pump current was increased and decreased in steps and fine-tuned for lowest RF linewidth, while monitoring the optical spectrum with an OSA for single-mode oscillation. Figure~\ref{fig:lorentzian_width} displays the Lorentzian linewidth component versus pump current $I_{\textrm{p}}$ expressed as the threshold factor $X=(I_{\textrm{p}}-I_{\textrm{p,th}})/I_{\textrm{p,th}}$, where $I_{\textrm{p,th}}$ is the threshold pump current of approximately 42~mA. The error bars express the uncertainty in fitting. A doubly logarithmic plot is chosen to facilitate comparison with the expected inverse power law dependence of the linewidth as straight line with negative unity slope. The red line is a least-square fit with fixed negative unity slope versus the inverse threshold factor, $\frac{1}{X}$, showing that the measured linewidth narrows approximately inversely with laser power as theoretically expected. The power narrowing data do not display a levelling-off, in spite of significant intensity build-up in the high-finesse ring resonators. The lowest linewidth obtained from power spectral density recordings (shown as round symbol for comparison) is in good agreement with the data obtained from Lorentz fitting. The linewidth limit of 40~Hz in Fig.~\ref{fig:noise_PSD} is the narrowest intrinsic linewidth ever reported for a hybrid or heterogeneously integrated diode laser. 

\section{Conclusions}
We have demonstrated a hybrid integrated and widely tunable single-frequency diode laser with an intrinsic linewidth as low as 40~Hz, a spectral coverage of more than 70~nm and a maximum fiber-coupled output of 23~mW. The narrow linewidth is achieved via feedback from a low-loss dielectric waveguide circuit that extends the laser cavity to a roundtrip optical length of $2L^{\textrm{(o)}}=0.5$~m, in combination with single-mode resolving filtering. Realizing such high-finesse filtering with cascaded microring resonators with essentially a single roundtrip through a long and low-loss feedback arm allows strong linewidth narrowing in the presence of significant laser cavity roundtrip losses. The tolerance to loss in this approach is important because semiconductor amplifiers are intrinsically lossy, such as also the mode transitions between different waveguide platforms in hybrid or heterogeneously integrated photonic circuits. Choosing dielectric feedback waveguides based on silicon nitride is important for avoiding nonlinear loss because GW/cm$^2$-level intensities readily occur in lasers with tens of mW output and high-finesse intracavity filtering. The approach demonstrated here is promising for further linewidth narrowing through stronger pumping, as no hard linewidth limit through nonlinear loss is apparent with dielectric feedback circuits. Although some promise lies in further extension of the cavity length, as the cavity photon lifetime is not yet fully saturated, in combination with tighter filtering, our analysis shows that further significant improvement requires reduction of the propagation loss in the feedback circuit. This route appears very feasible because silicon nitride waveguides can be fabricated with extremely low loss down to 0.045~dB/m ~\cite{bauters_2011OE}, while several meter long silicon nitride resonator circuits have been demonstrated with a spectral selectivity better than 100~MHz ~\cite{taddei_2018PTL}. These properties and options indicate the feasibility of Hertz-level integrated diode lasers on a chip. 

\section*{Funding}
This research was funded by the IOP Photonic Devices program of RVO (Rijksdienst voor Ondernemend Nederland), a division of the Ministry for Economic Affairs, The Netherlands and in part by the European Union's Horizon 2020 research and innovation programme under grant agreement 780502 (3PEAT).

\section*{Disclosures}
The authors declare no conflicts of interest.




\end{document}